\documentstyle[sprocl,axodraw]{article}
\def\Journal#1#2#3#4{{#1} {\bf #2}, #3 (#4)}

\def\ANP{\em Ann.\ Phys.}
\def\CMP{\em Commun.\ Math.\ Phys.}
\def\IJMPA{{\em Int.\ J.\ Mod.\ Phys.}\ A}
\def\JMP{\em J.\ Mod.\ Phys.}
\def\MPLA{{\em Mod.\ Phys.\ Lett.}\ A}

\def\NPA{{\em Nucl.\ Phys.}\ A}
\def\NPB{{\em Nucl.\ Phys.}\ B}
\def\PLB{{\em Phys.\ Lett.}\  B}
\def\PRL{\em Phys.\ Rev.\ Lett.}
\def\PRA{{\em Phys.\ Rev.}\ A}
\def\PRD{{\em Phys.\ Rev.}\ D}
\def\ZPC{{\em Z.\ Phys.}\ C}

\def\be{\begin{equation}}
\def\ee{\end{equation}}
\def\bea{\begin{eqnarray}}
\def\eea{\end{eqnarray}}

\begin{document}

\title{MIXING AND CP VIOLATION: STATUS AND PROSPECTS}

\author{A. PILAFTSIS }

\address{Rutherford Appleton Laboratory\\
         Chilton, Didcot, OX11 0QX, UK}

\maketitle\abstracts{
Brief overview of the present status on mixing and CP violation in kaons and
$B$ mesons is given by means of the unitarity triangle. Theoretical
predictions on $\epsilon'/\epsilon$ are confronted with experimental results.
The prospects of detecting CP violation at high-energy $p\bar{p}$, $pp$,
$e^-e^+$ and $\mu^-\mu^+$ colliders are discussed in resonant scatterings
involving top quarks and/or heavy scalars. The relevance of the latter for
baryogenesis is outlined.} 

\section{Introduction}
Understanding the origin of charge-conjugation and parity (CP) violation in 
the $K^0-\bar{K}^0$ system must be considered as an important task of modern
physics, which may eventually help to address the fundamental question
concerning the observed asymmetry between matter and anti-matter, the
so-called baryon asymmetry in the Universe (BAU). Much theoretical as well as
experimental effort has been put to explore the discrete symmetries of
time-reversal (T) invariance, CP conservation, and CPT invariance. Even though
T/CP is violated in kaons, CPT is still a good symmetry of our quantum world, 
which has been tested experimentally to a high accuracy. CPT conservation is a 
generic feature emanating from a consistent field-theoretical description of 
our nature. 

In this brief review, we present some of the highlights regarding the topic of
T, CP and CPT violation as follows. In Sect.\ \ref{sec:discr}, classical
tests of T invariance and CPT invariance are mentioned, and possibilities of how
to break CPT and CP within the context of field theories are discussed.
In Sect.\ \ref{sec:triangle}, results for the Cabbibo-Kobayashi-Maskawa (CKM)
mixing angles represented by the unitarity triangle are given. More attention
is paid to the direct CP-violating parameter $\varepsilon'/\varepsilon$ in 
Sect.\ \ref{sec:epsprime}. In Sect.\ \ref{sec:resCP}, the prospects of
observing CP violation in resonant top and/or Higgs scatterings at $p\bar{p}$,
$pp$, $e^+e^-$ and $\mu^+\mu^-$ colliders are analyzed. In addition, the
significance of this kind of CP-violating phenomena for the BAU is briefly
described. Sect.\ \ref{sec:summary} summarizes our current knowledge of CP
violation and mixing for the present and future.

\section{Discrete symmetries: T, CP, and CPT}\label{sec:discr}
In this section, we will discuss some classical tests of discrete symmetries. 
So, T violation in kaons may be studied via the Kabir's observable
\cite{kabir} 
\be
\label{Al}
A_l(t) \ = \ \frac{ |\langle \bar{K}^0(t)|K^0(0)\rangle|^2\,
-\, |\langle K^0(t)|\bar{K}^0(0)\rangle|^2}{|\langle 
\bar{K}^0(t)|K^0(0)\rangle|^2\, +\, |\langle K^0(t)|\bar{K}^0(0)\rangle|^2}\
=\ 4 \Re e \varepsilon_K\, ,
\ee
where $\varepsilon_K$ is that parameter known from the $K^0-\bar{K}^0$ mixing.
Given the fact that the rule $\Delta S=\Delta Q$ holds phenomenologically,
$A_l(t)$ is practically measured at CPLear by comparing the decay chain,
$p\bar{p}\to K^+ \pi^- \bar{K}^0$; $\bar{K}^0 \to K^0 \to \pi^- e^+ \nu_e$, to
that of the decay sequence, $p\bar{p}\to K^- \pi^+ K^0$; $K^0 \to \bar{K}^0
\to \pi^+ e^- \bar{\nu}_e$. As a result, at large times ($t\to ``\infty"$),
$A_l(t)$ modifies to \cite{Tan/Dal} 
\be
\label{al}
a_l(``\infty")\ =\ 4\Re e \varepsilon_K \, -\, 2\Re e y_l\, ,
\ee
where $\Re e y_l$ is a CPT-violating term, which was found to be unobservably
small at CPLear. Two remarks regarding $A_l(t)$ are now in order. First, a
non-vanishing value of $A_l(t)$ would signify T and CP violation
independently, without having to resort to the CPT theorem. In the
Weisskopf--Wigner (WW) approximation, $A_l(t)$ turns out to be a constant of
time. Even though this is a limitation of the WW approach as was already
noticed by Khalfin,\cite{chi/sud} the deviation from constancy is mainly
present at very short (the quantum Zeno region) or very long (power law
regime) times. These phenomena have been estimated to be very far beyond the
experimental feasibility.\cite{chi/sud} Their origin may be traced to the
unitarity of the quantum nature.\cite{kabir} 

It is now interesting to discuss various alternatives of how to break CPT
and what kind of experimental tests can be carried out to probe CPT invariance.
There are few ways to break CPT:
\begin{itemize}
\item An obvious attempt would be to provide different masses for particles
and anti-particles; this implies a non-Hermitean Hamiltonian. Anti-gravity
models,\cite{antigravity} which predict that anti-particles should experience
a very small repulsive force within gravitational fields, rely effectively on
this option. Since such theories violate the equivalence principle of
Einstein, a consistency check for a weaker version of it has been suggested at
CPLear.\cite{zioutas} 
\item Another option may be based on Hawking's observation on the spectrum of
radiation of black holes.\cite{Hawking} Hawking has demonstrated that a
generalized description of quantum mechanics including gravity allows the
evolution of pure states into mixed ones, thus leading to a dynamics that
violates conventional quantum mechanics and so breaks CPT. This idea has been
applied to $K^0-\bar{K}^0$ system by the authors in Ref.\cite{Ellisetal} 
\item The authors of Ref.\cite{Tod/Oks} formulated an infinite component field
theory which effectively steps outside the standard assumption that field
theories have to be local. Experiments probing locality with kaons at 
DA$\Phi$NE have already been proposed.\cite{Eber}
\end{itemize}
Note that no known example of a theory exists as yet, in which CPT is broken
spontaneously. The most crucial experiments testing the validity of CPT to a
high accuracy are those involving kaons. More explicitly, one has that 
\begin{itemize}
\item $M_{K^0}=M_{\bar{K}^0}$. Experiments give the upper bounds on
$M_{K^0}-M_{\bar{K}^0}/M_{K^0}$: $3.5\times 10^{-18}$ (NA31),\cite{NA31} 
$1.3\times 10^{-18}$ (E773) \cite{E773} and $1.8\times 10^{-18}$ 
(CPLear).\cite{CPLear} 
\item $\Delta\varphi=\varphi_{00}-\varphi_{+-}=0$, where $\varphi_{00}$ and
$\varphi_{+-}$ are the phases of the known amplitude ratios $\eta_{00}$ and
$\eta_{+-}$, respectively. On the experimental side, we have
$\Delta\varphi^{exp} = 0.2^\circ\pm 2.6^\circ\pm 1.2^\circ$ (NA31) \cite{NA31} 
and $0.62^\circ\pm 0.71^\circ\pm 0.75^\circ$ (E773).\cite{E773}
\item the theoretical value of the superweak phase $ \varphi_{sw} =
\arctan(-2\Delta M_K/\Gamma)$ $ = 43.37^\circ\pm 0.17^\circ$, if one assumes
that CP violation originates mainly from $K\to \pi\pi$, and one can hence make
explicit use of the Bell-Steinberger relation. Experiments are consistent with
this result so far. 
\end{itemize}
Even though breaking of CPT may thwart basic field-theoretical requirements, 
local gauge field theories admit CP violation in general. There are mainly
two avenues that achieve that purpose:
\begin{itemize}
\item CP violation is explicitly broken by introducing arbitrary complex
phases in the Yukawa couplings. For example, the CKM matrix of the SM owes his
origin to this mechanism. 
\item CP violation is broken spontaneously. In this case, the original
Lagrangian preserves CP but not the vacuum state. So, after spontaneous
symmetry breaking, CP-violating interactions are induced. This mechanism
naturally takes place in multi-Higgs scenarios, such as the two-Higgs doublet
model of T.D. Lee or Weinberg's three-Higgs doublet model. 
\end{itemize}
Another interesting possibility that exploits both ideas is based on the fact
that the CP invariance of the Higgs sector gets broken radiatively, through
the presence of very heavy particles, {\em e.g.}, through heavy Majorana
neutrinos.\cite{IKP}

\section{The unitarity triangle}\label{sec:triangle}
The most efficient way to encode all the information for the mixing
parameters is through the Wolfenstein parametrization of the CKM, {\em viz.} 
\be
\label{CKM}
V_{ij}\ =\ \left( \begin{array}{ccc}
1-\lambda^2/2 & \lambda & A\lambda^3( \rho - i \eta) \\
-\lambda & 1-\lambda^2/2 & A\lambda^2 \\
A\lambda^3 (1-\rho-i\eta) & -A\lambda^2 & 1 \end{array} \right).
\ee
In Eq.\ \ref{CKM}, $V_{ij}$ has been expanded up to the third order
of the Cabbibo angle $\lambda=|V_{us}|=0.2205\pm 0.0018$. In this
parametrization, $A=|V_{cb}|/|V^2_{us}|=0.80\pm 0.04$, whereas less
determined are the parameters $\rho$ and $\eta$, which are important
to describe CP violation in the SM. The unitarity of the CKM matrix
allows one to represent the constraints on the mixing angles graphically,
by means of a triangle in the complex plane. Among all the six possible 
unitarity relations, the most useful one is given by
\be
\label{triangle}
V^*_{ud}V_{ub}\, +\, V^*_{cd}V_{cb}\, +\, V^*_{td}V_{tb}\ =\ 0.
\ee 
If we now divide the lhs of Eq. \ref{triangle} by $V^*_{cd}V_{cb}$, the one 
side of the triangle will be normalized to unity, while its angles will remain
unaffected as is shown in Fig.\ 1. In the same figure, the various constraints
on the combined $\rho-\eta$ values are 
\begin{center}
\begin{picture}(300,250)(0,0)
\SetWidth{1.}
\LinAxis(49.9,50)(250.1,50)(6,2,3.,0,1.)\Text(15,200)[l]{$\bar{\eta}$}
\LinAxis(49.9,216.66)(250.1,216.66)(6,2,-3.,0,1.)
                                          \Text(230,30)[r]{$\bar{\rho}$}
\LinAxis(50,50)(50,216.66)(5,2,-3.,0,1.)
\LinAxis(250,50)(250,216.66)(5,2,3.,0,1.)

\Text(50,45)[t]{$-0.6$}\Text(83.33,45)[t]{$-0.4$}\Text(116.66,45)[t]{$-0.2$}
\Text(150,45)[t]{$0.0$}\Text(183.33,45)[t]{$0.2$}\Text(216.66,45)[t]{$0.4$}
\Text(250,45)[t]{$0.6$}

\Text(45,83.33)[r]{$0.2$} \Text(45,116.66)[r]{$0.4$} \Text(45,150)[r]{$0.6$}
\Text(45,183.33)[r]{$0.8$} \Text(45,216.66)[r]{$1.0$}

\Line(150,50)(316.66,50)\Line(150,50)(160,113)\Line(316.66,50)(160,113)
\Text(316.66,45)[t]{$1.0$}
\Text(155,55)[l]{\small $\gamma$} \Text(152,80)[r]{\small $S_\beta$}
\Text(294,55)[r]{\small $\beta$} \Text(220,95)[l]{\small $S_\gamma$}

\CArc(150,50)(73.33,0,180)\CArc(150,50)(43.33,0,180)
\Text(85,60)[l]{\small $V_{ub}$}
\DashCArc(316.66,50)(201.66,125,180){3.}
\DashCArc(316.66,50)(128.33,122,180){3.}\Text(210,180)[]{\small $B_d$ mixing}

\DashCurve{(50,80)(150,93.3)(250,120)}{1.}
\DashCurve{(50,100)(150,120)(250,153)}{1.}
\Text(55,90)[l]{\small $\varepsilon_K$}

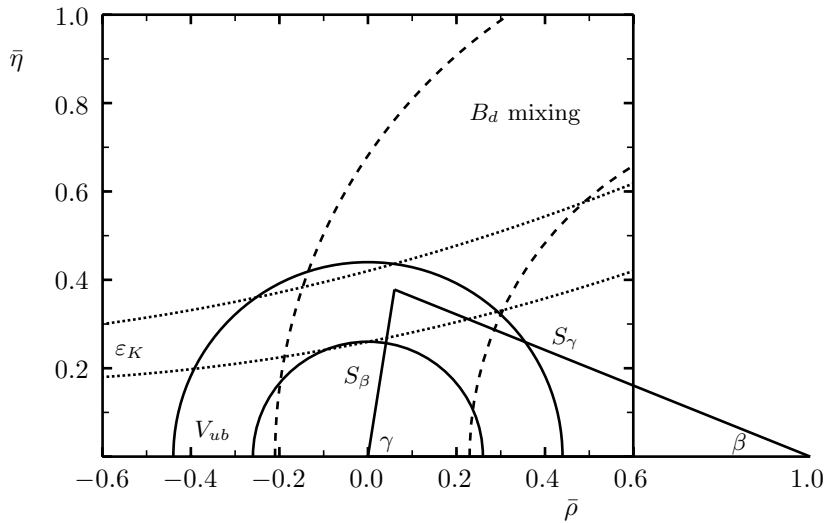
\Text(10,10)[l]{{\small {\bf Fig.\ 1:} Constraints on the unitarity triangle
($S_\beta=\frac{V^*_{ud}V_{ub}}{V^*_{cd}V_{cb}}$, 
$S_\gamma=\frac{V^*_{td}V_{tb}}{V^*_{cd}V_{cb}}$)}}
\end{picture}
\end{center}
\newpage
\noindent
implemented.  In particular, the length of $S_\beta$ is confined to lie
between the two semicircles determined by $|V_{ub}|/|V_{cb}| = 0.08\pm 0.03$,
coming from semi-leptonic $B$ meson decays. Similarly, the side $S_\gamma$ is
restricted by two arcs centered at $(\bar{\rho}, \bar{\eta}) = (1,0)$, which
are obtained from $B_d$ mixing effects. Finally, there are tight constraints
originating from indirect CP violation in $K^0$ 
mesons.\cite{DFP}$^-$\cite{Her/Nier}
From Fig.\ 1, it is worth noticing that much effort must be put to improve the
limits coming from $B$ physics. This also has been the scope of many recent
papers.\cite{AKL} 

\section{The status of $\varepsilon'/\varepsilon$}\label{sec:epsprime}
Of most theoretical as well as phenomenological importance is the question
concerning the actual value of the known direct CP-violating parameter
$\varepsilon'/\varepsilon $. Experimental results and theoretical
predictions cannot conclusively exclude any vanishing value for  
$\varepsilon'/\varepsilon $ so far. To be specific, the situation is
experimentally as follows:
\bea
\mbox{NA31:\cite{NA31'}}
                    &&\quad (23.0\, \pm\, 3.6\, \pm\, 5.4)\times 10^{-4}\, ,\\
\mbox{E731:\cite{E731'}} &&\quad (7.4\, \pm\, 5.2\, \pm\, 2.9)\times 10^{-4}.
\eea
Even though NA31 appears to rule out the superweak model, which predicts
CP violation in $\Delta S = 2$ transitions only, E731 is still consistent
with such a realization. On the theoretical side, there have been a number
of improvements that have been taken place over the last years,\cite{Fly/Ran}
including the top discovery which has enabled more accurate
renormalization-group-equation (RGE) studies. There are mainly three groups
working on this topic, using different approaches. Their results may be
summarized as follows: 
\bea
\mbox{I.\cite{CFMR}}  && (3.1\, \pm\, 2.5\, \pm\, 0.3)\times 10^{-4}\, ,\\
\mbox{II.\cite{BJL}}  && (6.7\, \pm\, 2.6)\times 10^{-4}\qquad 
                                                        (m_s=150\ \mbox{MeV}),\\
\mbox{III.\cite{HPSW}} && (9.9\, \pm\, 4.1)\times 10^{-4}\qquad 
                                                        (m_s=175\ \mbox{MeV}).
\eea
The errors in their estimates originate mainly from the different assumptions
made for the input data as well as from other uncertainties inherent to
the approach used. Yet, much theoretical improvement is needed to come, so
as to clarify the possibility of any beyond-the-SM CP violation.

\section{Resonant CP violation and the BAU}\label{sec:resCP}
It is now important to gauge our chances of finding CP violation at future
high-energy scatterings, which can take place at multi-TeV $pp$ or $p\bar{p}$
machines 
\newpage
\begin{center}
\begin{picture}(340,100)(0,0)
\SetWidth{1.}
\GOval(180,60)(30,30)(0){0.3} 
\DashArrowLine(110,60)(150,60){3.}\Text(130,70)[c]{$A,\ H$}
\DashArrowLine(210,60)(250,60){3.}\Text(230,70)[c]{$H,\ A$}
\ArrowLine(90,90)(110,60)\Text(87,85)[r]{$\mu^-_L$}
\ArrowLine(110,60)(90,30)\Text(87,35)[r]{$\mu^+_L$}
\ArrowLine(250,60)(270,90)\Text(273,85)[l]{$f$}
\ArrowLine(270,30)(250,60)\Text(273,35)[l]{$\bar{f}$}
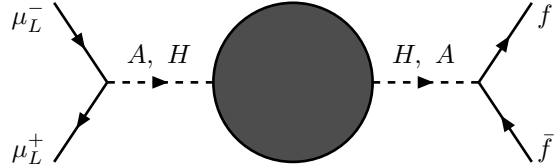
\Text(180,10)[]{{\small {\bf Fig.\ 2:} Resonant CP-violating $HA$ transitions}}
\end{picture}
\end{center}
({\em e.g.}\ LHC or TEVATRON), TeV-$e^-e^+$ colliders ({\em e.g.}\ NLC), or
$\mu^+\mu^-$ colliders with variable TeV energy.\cite{yuan} Particular
promising seem to be certain CP-violating observables, which can be resonantly
enhanced by particle widths.\cite{APres}$^-$\cite{mumures}
\vskip 11.cm
\centerline{{\small {\bf Fig.\ 3:} Production cross-section $\sigma$ 
(solid line) and $A_{CP}$ (dashed line)}}
\newpage
\noindent
In simple terms, the focal idea \cite{APres} may be explained as follows. At
high-energy processes, {\em e.g.}, $p(W^+)\bar{p}(d)\to (t^*, t'^*, \dots) \to
W^+ b$, heavy particles, such as the $t$ or $t'$ present in models with extra
quarks, can resonate, yielding a dynamical phase coming from the Breit-Wigner
propagator, 
\be
\label{BW}
\frac{1}{s-m^2+im\Gamma}.
\ee 
The imaginary CP-even phase, $-im\Gamma$, appearing in the transition
amplitude at $s\approx m^2$, will then be multiplied with the CP-odd phases of
the theory present in the interaction vertices, so as to produce a real
CP-violating contribution to the matrix element squared. Moreover, refinements
coming from CPT constraints \cite{CPBres,2HDMres} and gauge invariance
\cite{Pap/Pil} should be taken into consideration. In this context, another
important feature is that specific CP-violating observables based on
differential cross sections show a resonant behaviour as a function of
$s$.\cite{Pil/Now} We will elucidate this point in a $\mu^+\mu^-$
reaction.\cite{mumures} 

Recently, it has been argued \cite{mumures} that muon colliders is the 
most ideal place to search for CP-violating resonant transitions of a
CP-even Higgs scalar, $H$, 
into a CP-odd Higgs scalar, $A$, as shown in Fig.\ 2. Assuming that the
facility of longitudinal beam polarization is available, one can look for the
CP observable 
\be
\label{Acp}
A_{CP}\ =\ \frac{\sigma (\mu^-_L\mu^+_L\to f\bar{f}) - 
\sigma (\mu^-_R\mu^+_R\to f\bar{f})}{\sigma (\mu^-_L\mu^+_L\to f\bar{f}) +
\sigma (\mu^-_R\mu^+_R\to f\bar{f})}\, .
\ee
The $HA$ mixing can naturally be induced by heavy Majorana fermions.
As such, one may think either of heavy neutralinos and/or charginos in a 
supersymmetric SM or of heavy Majorana neutrinos present in E$_6$ motivated
models. Adopting the latter realization,\cite{mumures} we display our
numerical estimates of this analysis in Fig.\ 3. Notice that the mechanism 
of resonant CP violation is quite important to render $A_{CP}$ measurable.

C and CP violation is also one of the three Sakharov's necessary conditions
for generating the BAU, together with B violation and the requirement that the 
interactions should be out of thermal equilibrium. In general, there are two
known scenarios for baryogenesis. In the first scenario, the BAU is generated
at the electroweak phase transition,\cite{KRS} through instanton-type objects
(sphalerons) which violate B.\cite{manton} In the SM,  the so-generated BAU
appears to be small.\cite{Far/Shap,Gavelaetal} However, this is not true in
new-physics CP-violating scenarios.\cite{CKN} In the second scenario, 
baryogenesis is produced via B-violating decays of heavy particles in the
context of grand unified models, such as SO(10). Using the terminology known
from kaons, one can differentiate between two mechanisms of CP violation:
(i) CP violation present in the decay amplitudes (or $\varepsilon'$-type
effects), (ii) CP violation occuring in the mass matrix (or $\varepsilon$-type
effects).\cite{KRS} The latter may be related to the resonant CP-violating 
mechanism mentioned above, even though the situation is slightly different in
scatterings due to additional interference amplitudes. Finally, one could
exploit the fact that sphalerons violate B$+$L to convert an excess in L
into the observed excess in B. This can be achieved by L-violating
decays of heavy Majorana neutrinos,\cite{Fuk/Yan} which possess both kinds
of CP-violating interactions discussed above, {\em i.e.}, $\varepsilon$ and 
$\varepsilon'$-type.\cite{FPS}

\section{Summary: present and future}\label{sec:summary}
In summary, the present status on T/CP/CPT violation may be described as
follows:
\begin{itemize}
\item Despite the many experimental searches, CPT is still a good symmetry
of nature.
\item The origin of CP violation is not yet known in the $K^0$ system.
In fact, it must be specified whether CP non-conservation arises due to
the CKM matrix or CP is broken spontaneously, or there is another novel 
origin.
\item The knowledge of the top mass and the resulting improved RGE analyses 
for the  $K^0-\bar{K}^0$ mixing have given rise to more accurate theoretical
predictions for $\varepsilon_K$, thus leading to tighter constraints
on the $\rho-\eta$ plane.
\item The experimental as well as theoretical situation of 
$\varepsilon'/\varepsilon$ still remains inconclusive. In particular, we
do not know yet whether $\varepsilon'/\varepsilon \not= 0$ or whether CP 
violation occurs in $\Delta S=2$ transitions only.  
\end{itemize}
As for the future prospects of testing CP, many options are open. Perhaps, the
most appealing ones are given below. 
\begin{itemize}
\item Many tests of CP violation with $B$ mesons are performed or planned 
to take place in the so-called $B$-meson factories, {\em e.g.},
KEK, SLAC, HERA-B, {\em  etc}. Such tests will probe the sum of all angles
in the unitarity triangle with a good precision and may hence consistently
check if the CKM-mixing matrix can adequately describe low-energy CP violation.
\item Reducing the uncertainties of $\varepsilon'/\varepsilon $ below the 
bench-mark of $3.\ 10^{-4}$ will be one of the primary aims of future 
experiments, {\em e.g.}, DA$\Phi$NE.
\item At high-energy colliders, there are several CP-violating observables
based on the top or Higgs production and decay, which are resonantly enhanced
and are very sensitive to new-physics CP-violating scenarios. Detecting
CP-violating phenomena at resonant high-energy scatterings will give another
viewpoint of understanding the observed BAU, which also calls for new-physics
CP violation.
\item There is need for independent tests of T violation, such as looking at
possible electric dipole moments for $n$, $e$, $\mu$, $\tau$ and/or
electric dipole form factors of $b$ and $t$.\cite{EDMs,Nelson}
\end{itemize}

\end{document}